\documentclass[10pt,twocolumn,english,aps,manuscript,showpacs]{revtex4}
\usepackage[T1]{fontenc}
\usepackage[latin1]{inputenc}
\usepackage{geometry}
\usepackage{babel}

\makeatletter

\providecommand{\LyX}{L\kern-.1667em\lower.25em\hbox{Y}\kern-.125emX\@}

\makeatother
\begin{document}

\pacs{05.40.-a, 02.50.Ey, 05.60.-k}

\title{Breakdown of the Fluctuation-Dissipation Theorem for fast superdiffusion}

\author{Ismael V. L. Costa, Rafael Morgado, Marcos V. B. T. Lima and Fernando
A. Oliveira}

\address{Institute of Physics and International Center of Condensed Matter
Physics, University of Bras\'{\i}lia, CP 04513, 70919-970, Bras\'{\i}lia-DF,
Brazil}

\date{\today {}}

\begin{abstract}
We study anomalous diffusion for one-dimensional systems described
by a generalized Langevin equation. We show that superdiffusion can
be classified in slow superdiffusion and fast superdiffusion. For
fast superdiffusion we prove that the Fluctuation-Dissipation Theorem
does not hold. We show as well that the asymptotic behavior of the
response function is a stretched exponential for anomalous diffusion
and an exponential only for normal diffusion.
\end{abstract}
\maketitle
Since its formulation, the Fluctuation-Dissipation Theorem (FDT) has
played a central role\cite{Kubo, kubo2} in non-equilibrium statistical
mechanics (NESM). It reaches such an importance that a full formulation
of NESM is given \cite{kubo2} based on it. In the last 30 years,
fundamental concepts and methods have been developed \cite{Kubo}-\cite{Lee3}
and a large number of connections have been established (see ref.
\cite{Lee2} and references therein). A necessary requirement for
the FDT is that the time-dependent dynamical variables are well defined
at equilibrium. The presence of far from equilibrium dynamics may
lead to situations where the FDT does not hold, the aging process
in spin-glass systems being a good example \cite{Parisi}-\cite{Grigera}. 

Diffusion is one of the simplest processes by which a system reaches
equilibrium. For normal diffusion, the process is so well known that
it may be described by an equilibrium type distribution for the velocity
and position of a particle. However, the strange kinetics of anomalous
diffusion, intensively investigated in the last years \cite{Bouc}-\cite{Oliveira},
shows surprising results. Consequently, studying anomalous diffusion
seems to be the best way to obtain the conditions of validity for
the FDT. 

In this letter, we present a straightforward proof of the inconsistency
of the FDT for a certain class of superdiffusive processes described
by a generalized Langevin equation (GLE). The use of the FDT allows
us to classify two classes of superdiffusion. The first class, which
we shall call slow superdiffusion, does obey the FDT; the second class,
which we shall call fast superdiffusion, does not obey the FDT. The
proof is simple and we discuss as well how the diffusive process leads
to an equilibrium. 

We shall start writing the GLE for an operator \( A \) in the form
\cite{Kubo, Mori, Lee2}

\begin{equation}
\label{1}
\frac{dA(t)}{dt}=-\int _{0}^{t}\Gamma (t-t')A(t')dt'+F(t),
\end{equation}
 where \( F(t) \) is a stochastic noise subject to the conditions
\( \langle F(t)\rangle =0 \), \( \langle F(t)A(0)\rangle =0 \) and

\begin{equation}
\label{2}
C_{F}(t)=<F(t)F(0)>=<A^{2}>_{eq}\Gamma (t).
\end{equation}
 Here \( C_{F}(t) \) is the correlation function for \( F(t) \)
and the brackets \( <> \) indicate thermal average. Eq. (\ref{2})
is the famous Kubo FDT and it is quite general. In principle, the
presence of the kernel \( \Gamma (t) \) allows us to study a large
number of correlated processes. 

We may naively expect that, by Eq. (\ref{1}) and Eq. (\ref{2}),
a system will be driven to an equilibrium , i.e. 

\begin{equation}
\label{2.6}
\lim _{t\rightarrow \infty }<A^{2}(t)>=<A^{2}>_{eq}.
\end{equation}
 We shall see however that this is not always the case for superdiffusive
dynamics. Let us define the variable 

\begin{equation}
\label{3}
y(t)=\int _{0}^{t}A(t')dt',
\end{equation}
 with asymptotic behavior 

\begin{equation}
\label{3.5}
\lim _{t\rightarrow \infty }<y^{2}(t)>\sim t^{\mu }.
\end{equation}
 For normal diffusion \( \mu =1 \), we have subdiffusion for \( \mu <1 \)
and superdiffusion for \( \mu >1 \). Notice that if \( A(t) \) is
the momentum of a particle with unit mass, \( y(t) \) is its position.
Using Kubo's definition of the diffusion constant we get \cite{Oliveira}

\begin{equation}
\label{4}
D=\lim _{z\rightarrow 0}\frac{<A^{2}>_{eq}}{\tilde{\Gamma }(z)},
\end{equation}
 where \( \tilde{\Gamma }(z) \) is the Laplace transform of \( \Gamma (t) \).
A finite value of \( \tilde{\Gamma }(0) \) corresponds to normal
diffusion, \( \tilde{\Gamma }(0)=0 \) to superdiffusion and \( \tilde{\Gamma }(0)=\infty  \)
to subdiffusion. Notice that 

\begin{equation}
\label{5}
\gamma =\tilde{\Gamma }(0)=\int _{0}^{\infty }\Gamma (t)dt
\end{equation}
 plays the same role as the friction in the usual Langevin's equation,
i.e., GLE without memory. 

Now we propose a solution for Eq. (\ref{1}) as 

\begin{equation}
\label{6}
A(t)=\int _{0}^{t}R(t-t')F(t')dt',
\end{equation}
 where we have set \( A(0)=0 \) and \cite{Srokowski1}

\begin{equation}
\label{7}
\tilde{R}(z)=\frac{1}{z+\tilde{\Gamma }(z)}.
\end{equation}
 Squaring Eq. (\ref{6}) and taking thermal average we obtain 

\begin{equation}
\label{8}
<A^{2}(t)>=\int _{0}^{t}\int _{0}^{t}C_{F}(t'-t'')R(t')R(t'')dt'dt''.
\end{equation}

At this point, it is quite usual to perform numerical calculation
\cite{Srokowski1}. From Eq. (\ref{1}), we can get a self-consistent
equation for \( R(t) \) as 

\begin{equation}
\label{8.5}
\frac{dR(t)}{dt}=-\int _{0}^{t}\Gamma (t-t')R(t')dt'.
\end{equation}
By using the FDT Eq.(\ref{2}) and Eq.(\ref{8.5}) we can exactly
integrate Eq. (\ref{8}) and obtain 

\begin{equation}
\label{9}
<A^{2}(t)>=<A^{2}>_{eq}\lambda (t),
\end{equation}
 where 

\begin{equation}
\label{10}
\lambda (t)=1-R^{2}(t).
\end{equation}
 Notice now that Eq. (\ref{2.6}) is satisfied if and only if 

\begin{equation}
\label{10.5}
\lim _{t\rightarrow \infty }\lambda (t)=\lambda ^{*}=1,
\end{equation}
 or equivalently 

\begin{equation}
\label{11}
\lim _{t\rightarrow \infty }R(t)=\lim _{z\rightarrow 0}z\tilde{R}(z)=0.
\end{equation}
 Equation (\ref{11}) is the ergodic condition \cite{Lee3}. It is
satisfied for normal diffusion and subdiffusion. Now for superdiffusive
systems 

\begin{equation}
\label{12}
\lim _{t\rightarrow \infty }R(t)=(1+b)^{-1},
\end{equation}
 where 

\begin{equation}
\label{12.5}
b=\lim _{z\rightarrow 0}\frac{\partial \tilde{\Gamma }(z)}{\partial z}.
\end{equation}
 There are two distinct limits for \( b \), which define two classes
of superdiffusion. For the first class, \( b=\infty  \) and the system
obeys the FDT. The second class has \( b\neq \infty  \) and does
violate the FDT. The first class we shall call slow superdiffusion
(SSD) and the second class fast superdiffusion (FSD). 

Consider now the asymptotic behavior for \( \tilde{\Gamma }(z) \)

\begin{equation}
\label{13}
\lim _{z\rightarrow 0}\tilde{\Gamma }(z)=az^{\nu -1}.
\end{equation}
 For \( \nu <1 \) we have subdiffusion, for \( \nu =1 \) normal
diffusion. For \( 1<\nu <2 \) the process belongs to the SSD and,
finally, for \( \nu \geq 2 \) we have FSD. There is an obvious connection
between \( \nu  \) and \( \mu  \). Using Eq. (\ref{3.5}) and the
fact that \( \lim _{z\rightarrow 0}\tilde{\Gamma }(z)=\lim _{t\rightarrow \infty }\tilde{\Gamma }(1/t) \)
we get \( \nu =\mu  \) and consequently the FSD starts at \( \mu \geq 2 \),
i.e., the ballistic motion and beyond. It is interesting to note that
Lee \cite{Lee3} proved the failure of ergodicity for the ballistic
motion and now we showed that the FDT does not hold for this motion.

Now we test our analysis against simulations. Let us consider the
function 

\begin{equation}
\label{14}
\Gamma (t)=\beta \left[ \frac{\sin (w_{2}t)}{t}-\frac{\sin (w_{1}t)}{t}\right] ,
\end{equation}
 where \( w_{2}>w_{1} \). This function was chosen so that \( \tilde{\Gamma }(0)=0 \)
for any \( w_{1}\neq 0 \) . Thus, for \( w_{1}=0 \) we have normal
diffusion and for any \( w_{1}\neq 0 \) we have superdiffusion with
\( \mu =2 \). If we let \( \beta =w_{2}/2 \) we get \( \lambda ^{*} \)
as 

\begin{equation}
\label{15}
\lambda ^{*}=1-\left( \frac{2w_{1}}{w_{1}+w_{2}}\right) ^{2}.
\end{equation}
 Any value of \( \lambda ^{*} \) different from \( 1 \) shows the
inconsistency of the FDT in Eq. (\ref{2}), because we start supposing
the existence of an equilibrium value \( <A^{2}>_{eq} \) and, after
an infinite time, we end up with \( <A^{2}>_{eq}\lambda ^{*} \).
No matter the \( <A^{2}>_{eq} \) that we input in Eq. (\ref{2}),
we never reach it, except for the trivial null value.

\begin{figure}

\caption{Normalized mean square velocity as a function of time for the memory
given by Eq.(\ref{14}). Here \protect\( \beta =w_{2}/2\protect \)
and \protect\( w_{2}=0.5\protect \). Each curve corresponds to a
different value of \protect\( w_{1}\protect \). a) \protect\( w_{1}=0\protect \);
b) \protect\( w_{1}=0.25\protect \); c) \protect\( w_{1}=0.45\protect \).
The horizontal lines correspond to the final average value \protect\( \lambda _{s}\protect \).
In agreement with the theoretical prediction, \protect\( \lambda _{s}\protect \)
decreases as \protect\( w_{1}\protect \) grows.}
\end{figure}

Now we select \( A(t)=v(t) \), the particle's velocity, so that \( <v^{2}(t)>=<v^{2}>_{eq}\lambda (t) \).
We simulate the GLE for a set of \( 10,000 \) particles starting
at rest at the origin and using the memory in Eq. (\ref{14}) with
\( w_{2}=0.5 \) and different values of \( w_{1} \). The results
of these simulations are shown in Fig. 1, where we plot \( <v^{2}(t)> \).
We used the normalization \( <v^{2}>_{eq}=1 \), so that \( <v^{2}(t)>=\lambda (t) \).
Notice that \( \lambda (t) \) does not reach a stationary value,
rather it oscillates around a final average value \( \lambda _{s} \).
This value of \( \lambda _{s} \) should be compared with \( \lambda ^{*} \)
obtained from Eq. (\ref{15}). 

In Fig. 2 we plot \( \lambda ^{*} \) as a function of \( w_{1} \)
as in Eq.(\ref{15}) with \( w_{2}=0.5 \). We also plot the final
average values \( \lambda _{s} \) obtained from simulations for different
values of \( w_{1} \). Note that simulations agree with theory and
\( \lambda _{s}\rightarrow 1 \) when \( w_{1}\rightarrow 0 \) .

\begin{figure}

\caption{\protect\( \lambda ^{*}\protect \) as a function of the parameter
\protect\( w_{1}\protect \). Each dot corresponds to a value of \protect\( \lambda _{s}\protect \)
obtained from simulations like those described in Fig. 1. The line
corresponds to the theoretical prediction given by Eq.(\ref{15}).}
\end{figure}

For \( \mu >2 \), the FSD cannot be described by the methods we discussed
here. Once the FDT does not work, the GLE and the FDT together predict
results such as null dispersion for the dynamical variable, i.e. \( <A^{2}(t\rightarrow \infty )>=0. \)
Moreover, the exponent \( \mu  \) can be put as \( \mu =2/D_{F} \),
where \( D_{F} \) is the fractal dimension \cite{Ord}. Consequently
\( \mu >2 \) leads to \( D_{F}<1 \), which is not a full curve,
but a set of points such as the Cantor set, and cannot represent a
classical trajectory.

At first sight, the results presented here seem strange. Why does
the FDT not work for the FSD? As we remarked before, \( \gamma  \)
in Eq. (\ref{5}) plays the same role as the usual friction in the
Langevin Equation that yields \( R(t)\sim \exp (-\gamma t) \) with
a relaxation time \( \tau =\gamma ^{-1} \) for large times. For both
SSD and FSD, \( \gamma ^{-1}=\infty  \) and the system should not
reach an equilibrium. 

Now we address the previous question in another way: ''Why does the
FDT work for the SSD? Is it really \( \tau =\Gamma (0)^{-1} \) the
relaxation time?''. In order to answer this question one needs to
know the asymptotic behavior of \( R(t) \) as \( t\rightarrow \infty  \).
From Eq. (\ref{8.5}) we may write

\begin{equation}
\label{16.5}
\ln R(t)=-\Gamma (t)\int _{0}^{t}R(t')dt'-t\tilde{\Gamma }(z).
\end{equation}

In the limit when \( t\rightarrow  \)\( \infty  \) or, equivalently,
\( z=1/t\rightarrow 0 \), it is possible to eliminate the first term
at the right of Eq. (\ref{16.5}) by using

\begin{equation}
\label{16.8}
I=\lim _{t\rightarrow \infty }\Gamma (t)\int _{0}^{t}R(t')dt'=\lim _{z\rightarrow 0}\frac{z\tilde{\Gamma }(z)}{z+\tilde{\Gamma }(z)}.
\end{equation}

Notice that for \( \tilde{\Gamma }(z)=az^{\mu -1} \) and \( \mu >0 \),
\( I\rightarrow 0 \) and we get the asymptotic behavior 

\begin{equation}
\label{17}
\ln R(t)=-t\int _{0}^{t}\Gamma (t')dt'=-t\tilde{\Gamma }(0).
\end{equation}

The limit in Eq. (\ref{17}) is quite clear for normal diffusion,
where \( \gamma =\tilde{\Gamma }(0) \) is finite, and for subdiffusion,
where \( \tilde{\Gamma }(0)\rightarrow \infty  \). However, for superdiffusion
one must look carefully since \( \tilde{\Gamma }(0)\rightarrow 0 \).
We use \( \tilde{\Gamma }(z) \) as in Eq (\ref{13}) to obtain 

\begin{equation}
\label{18}
\lim _{t\rightarrow \infty }t\tilde{\Gamma }(0)=t\tilde{\Gamma }(1/t)=at^{2-\mu }.
\end{equation}
 We see that Eqs. (\ref{17}) and (\ref{18}) yield \( R(t\rightarrow \infty )=0 \)
only for \( \mu <2 \), what includes the subdiffusion, the normal
diffusion and the SSD. For the FSD, \( \mu \geq 2 \) and we shall
use Eq. (\ref{12}) to obtain the infinite limit. Thus, in this limit
process, there is an infinite relaxation time \( \tau =\gamma ^{-1} \)
for superdiffusion. However, this relaxation time can be seen only
as a result of an evolution, which, for the SSD, is never of the same
order of \( t \) in the limit \( t\rightarrow \infty  \). Consequently,
for long times, the SSD presents a finite relaxation time. In short,
the SSD has in common with normal diffusion and subdiffusion the fact
that they have a finite relaxation time and obey the FDT.  

Now we can look beyond the exponential aspect of the asymptotic solution
Eq. (\ref{17}) and use Eq. (\ref{18}) to obtain

\begin{equation}
\label{19}
\lim _{t\rightarrow \infty }R(t)=\exp \left[ -\left( \frac{t}{\tau }\right) ^{\beta }\right] ,
\end{equation}
 where 

\begin{equation}
\label{19.5}
\beta =2-\mu .
\end{equation}
 For \( \mu \neq 1 \), \( \tau =a^{-1/\beta } \) and for \( \mu =1 \),
\( \tau =\gamma ^{-1}=\tilde{\Gamma }(0)^{-1} \). The function Eq.
(\ref{19}) is a stretched exponential and we shall discuss that in
detail below.

We have important results. First, we obtain a stretched exponential
associated with anomalous diffusion, i. e. both subdiffusion and SSD.
Also, we obtain the exponent \( \beta  \) directly, not by fitting
nor simulations, with no reference to a specific system. Finally,
we show that the relaxation time of the correlation function is \( \tilde{\Gamma }(0)^{-1} \)
only for normal diffusion. For that case, the correlation function
decays as an exponential. For subdiffusion and for SSD the relaxation
time is associated with the coefficient of the main term of \( \tilde{\Gamma }(z) \)
in the limit when \( z\rightarrow 0 \). Thus we can define a relaxation
time for both normal and anomalous diffusion in the form

\begin{equation}
\label{20}
\tau =\lim _{z\rightarrow 0}\left[ z^{1-\mu }\tilde{\Gamma }(z)\right] ^{-\frac{1}{\beta }}.
\end{equation}

Notice that for \( \mu =\beta =1 \), \( \tau =\tilde{\Gamma }(0)^{-1} \)
as expected for normal diffusion.

Let us discuss the very particular behavior of \( \mu =0 \), i.e.
the {}``no diffusion at all'' behavior. This can be easily obtained
by the constant memory \( \Gamma (t)=\omega _{0}^{2} \), which yields
for the friction force in Eq. (\ref{1}) \( -m\omega _{0}^{2}y \).
This is precisely an harmonic oscillator, which does not dissipate
nor diffuse at all. For this system, we have \( \tilde{\Gamma }(z)=\omega _{0}^{2}z^{-1} \),
and \( R(t) \) can be exactly solved as a \( cos(\omega _{0}t) \)
type behavior. As expected, \( R(t) \) has no relaxation time. However,
using \( \tilde{\Gamma }(z) \) on Eq.(\ref{20}), we get \( \tau =\omega _{0}^{-1}, \)
which is the time scale of the oscillation, i.e. the inverse of the
frequency. Consequently, even in an extreme situation where we do
not have a relaxation time, Eq. (\ref{20}) yields the right time
scale of the system.

The research on the striking universality properties of slow relaxation
dynamics in glass \cite{Parisi, Xia}, supercooled liquids \cite{Xia},
liquid crystal polymer \cite{Benmouna} and disordered vortex lattice
in superconductors \cite{Bouchad2} has been driving great efforts
in the last decades. A large and growing literature can been found
where the non-exponential behavior (stretched exponentials) has been
observed in correlation functions \cite{Xia, Bouchad2}. Those have
in common the fact that they are subject to an anomalous diffusion.
Peyrard \cite{Peyrard} made a model for two-dimensional water and,
by using Monte Carlo simulation, obtained the correlation function
with an exponent \( 0.3<\beta <0.6 \). When the temperature decreases,
he suggests that \( \beta \rightarrow 1 \). Using his data in Eq.(\ref{19.5}),
we get \( \beta \sim 0.75 \). It would be too naive to expect that
our simple unidimensional, linear approach would describe all the
range of complex structures. Nevertheless, it may bring an insight
to guide us in such situations. 

In conclusion, we discussed the stationary behavior for the mean square
value of a dynamical variable \( A(t) \) and noticed that the superdiffusive
motion must be classified in slow superdiffusive (SSD) and fast superdiffusive
(FSD). The FSD motion shows an inconsistency between the GLE and the
FDT. The FSD has infinite relaxation time, and consequently never
reaches equilibrium. This kind of superdiffusion in which \( <A^{2}(t)>\sim t^{\mu } \)
with \( \mu \geq 2 \) is common in hydrodynamical processes. It is
not surprising that these processes will be far from equilibrium and
violate the FDT. We pointed out here how it happens and precisely
where the FDT breaks down. As we have already mentioned, spin glasses
seem to be a rich field for studying these phenomena. Indeed experimental
\cite{Grigera} and theoretical works \cite{Parisi, Ricci} have been
reported in this area, confirming the violation of the FDT. As well,
the stretched exponential behavior found in noncrystaline material
is connected here with anomalous diffusion. It would be very helpful
if the exponent \( \mu  \) for those diffusive processes could be
measured. Another related phenomenon is the anomalous reaction rate,
which we expect to discuss soon. Although anomalous diffusion remains
as a surprising phenomena, we hope that this work will help in the
centennial effort to understand diffusion and the relation between
fluctuation and dissipation. A generalization of the FDT to include
the FSD is necessary, what will require a deeper understanding of
systems far from equilibrium.

This work was supported by CAPES and CNPq - CTPETRO.

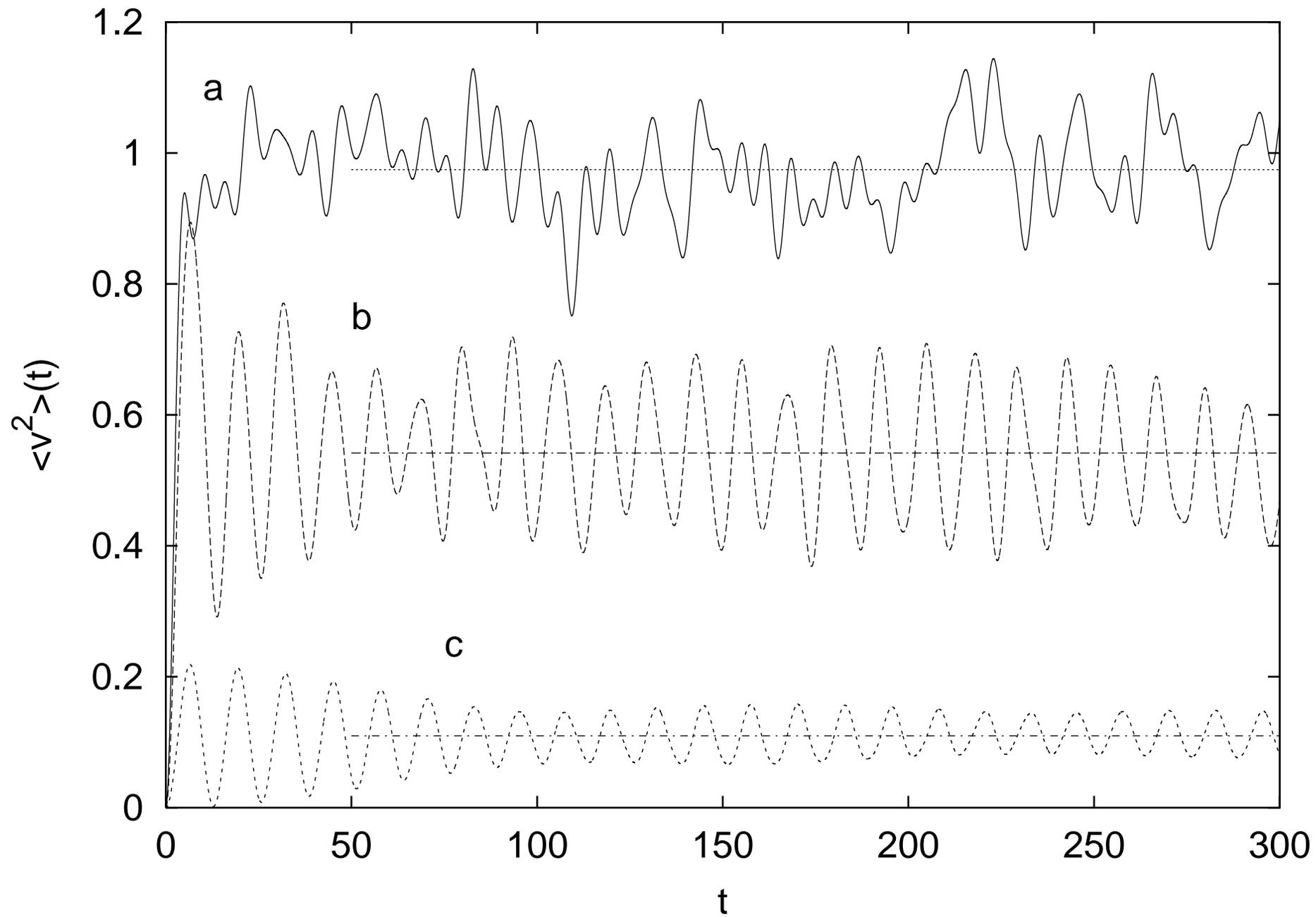
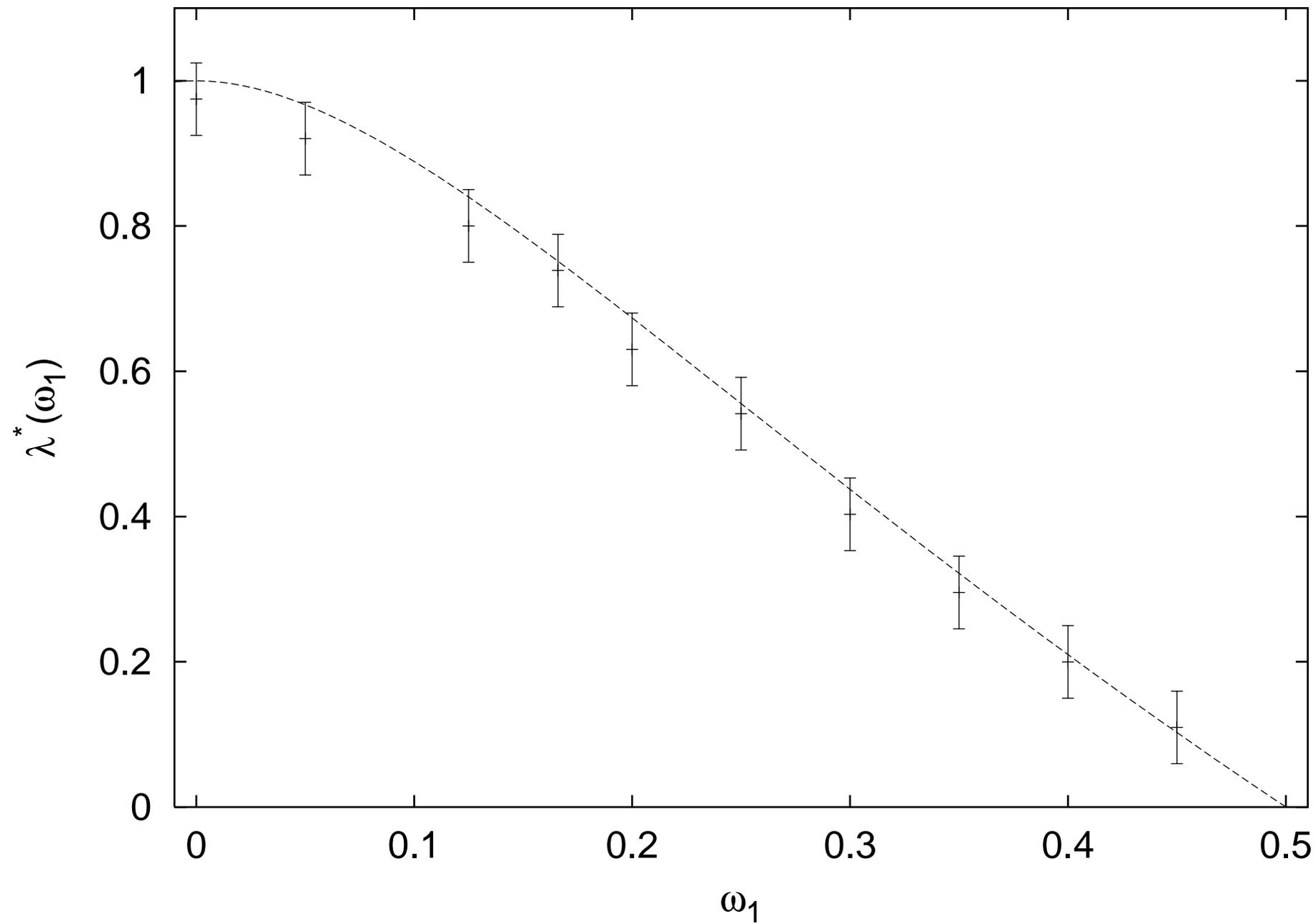
\end{document}